\documentclass[preprint,preprintnumbers,byrevtex,prb]{revtex4}
\usepackage{graphicx}

\begin{document}
\newcommand{\ethaffil}{Laboratory of Physical Chemistry, ETH Zurich, CH-8093 Zurich, Switzerland}
\newcommand{\humbaffil}{Nano-Optics, Humboldt University, Hausvogteiplatz 5-7, D-10117 Berlin,
Germany}

\title{Controlled photon transfer between two individual nanoemitters \\via shared
high-Q modes of a microsphere resonator}

\author{S.~G\"{o}tzinger}
\altaffiliation[Present address: ]{Edward L. Ginzton Laboratory,
Stanford University, Stanford, CA 94305, USA}
\affiliation{\humbaffil}
\author{L.~de S. Menezes}
\altaffiliation[Permanent address: ]{Depto. de F\'{\i}sica,
Universidade Federal de Pernambuco, 50670-901 Recife-PE, Brazil.}
\affiliation{\humbaffil}
\author{A.~Mazzei}
\affiliation{\humbaffil}
\author{S.~K\"uhn}
\affiliation{\ethaffil}
\author{V.~Sandoghdar}\email{vahid.sandoghdar@ethz.ch}
\affiliation{\ethaffil}
\author{O.~Benson}\email{oliver.benson@physik.hu-berlin.de}
\affiliation{\humbaffil}


\begin{abstract}
We realize controlled cavity-mediated photon transfer between two
single nanoparticles over a distance of several tens of
micrometers. First, we show how a single nanoscopic emitter
attached to a near-field probe can be coupled to high-Q
whispering-gallery modes of a silica microsphere at will. Then we
demonstrate transfer of energy between this and a second
nanoparticle deposited on the sphere surface. We estimate the
photon transfer efficiency to be about six orders of magnitude
higher than that via free space propagation at comparable
separations.

\end{abstract}

\maketitle

\bigskip

If two dipolar emitters are separated by a distance $r$ much less
than the transition wavelength $\lambda$, they can undergo strong
coherent dipole-dipole coupling, leading to sub- and
superradiance~\cite{Dicke:54,Hettich:02}. If their transitions are
broadened, dipole-dipole coupling becomes incoherent as in the
case of Fluorescence Resonant Energy Transfer (FRET), where the
energy from a ``donor" is transferred to an ``acceptor", provided
there is sufficient overlap between the former's emission spectrum
and the latter's absorption line. The efficiency of
FRET~\cite{Foerster} is proportional to $(1+(r/r_{0})^6)^{-1}$ and
falls to $50\%$ already at $r=r_{0} \sim 10~nm$. For $r>\lambda$,
optical communication between the two emitters takes place via
propagating photons, while the coupling drops as $1/r^2$. At a
distance of $50~\mu m$, the efficiency of one emitter absorbing a
photon radiated by the other is merely $3\times 10^{-13}$,
considering a typical room temperature absorption cross
section~\cite{Schaefer} of $\sigma_A \approx 10^{-16}\, cm^2$.

In order to enhance the coupling between two emitters, one could
funnel the energy from one to the other by using optical elements
such as lenses and waveguides. However, this process remains
limited because each photon flies by the atom only once. Thus, it
is interesting to exploit resonant structures to provide longer
effective interaction times. In addition, resonators can influence
radiative processes by modifying the density of states
\cite{Berman_book_1996,Chang_Campillo}. Furthermore, if the cavity
is made very small, the field per photon becomes large, resulting
in a much stronger coupling between the emitter and the photon
field. These effects depend on the three-dimensional locations of
the donor and acceptor molecules in a decisive manner. Enhancement
of the energy transfer rate in microcavities has been previously
demonstrated for systems containing ensembles of molecules
distributed over a volume (area) much larger than $\lambda^3$
($\lambda^2$) in microdroplets~\cite{Folan_PRL_1985,ArnoldJCP96}
or polymer microcavities~\cite{HopmeierPRL99}. However, the ideal
case where two single emitters couple via photon transfer through
a single mode of a high finesse microcavity remains a great
experimental challenge. In this Letter we report on a major step
toward this goal. We present experimental results on the
controlled optical coupling and photon transfer between two
individual subwavelength emitters at large distances via high-Q
modes of a microresonator.

\begin{figure}[b!] \centering
\includegraphics[width=10 cm]{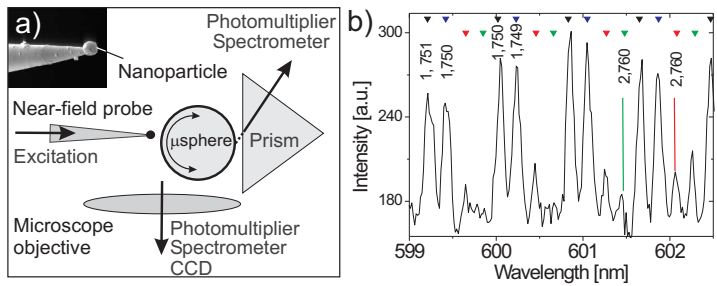}
\caption{a) The schematics of the experimental setup. The inset
shows an SEM image of a single $500~nm$ bead attached to a glass
tip. b) Spectrum recorded via the prism when the bead was close to
the sphere's surface. Colored triangles indicate the theoretical
positions of the resonances assuming a sphere of diameter 96 $\mu
m$ and an index of refraction of N=1.45724. The fundamental modes
are marked in  black (TE) and blue (TM). Modes with $n=2$ are
marked in red (TE) and green (TM). The first label denotes $n$ and
the second corresponds to the \emph{l-}number.} \label{fig1}
\end{figure}

Silica microspheres melted at the end of a fiber support very
high-Q Mie modes known as whispering-gallery modes
(WGMs)~\cite{Chang_Campillo}. WGMs are characterized by the radial
number $n$ and angular numbers $l$ and $m$ which determine their
resonance frequencies and spatial intensity
distributions~\cite{Chang_Campillo}. Each mode can have
polarizations TE or TM corresponding to radial magnetic and
electric fields, respectively. For a sphere of radius $R$ and
refractive index $N$, an increment in $l$ shifts the spectrum by
one free spectral range $FSR=c/(2\pi R N)$. The modes with $n=1$
and $l=\left\vert m \right\vert$ are called the ``fundamental"
modes and yield the most confined WGM with the largest electric
field at the sphere surface. The quantities $n$ and $l-\left\vert
m \right\vert+1$ give the number of intensity maxima along the
sphere radius and perpendicular to the equator,
respectively~\cite{KnightOptLett95}.

We first discuss the realization of an on-command coupling between
a dye-doped nanoparticle and the WGMs of a microsphere. As
depicted in Fig.~\ref{fig1}a, our strategy has been to attach the
nanoparticle to the end of a fiber tip (see inset) and use a
home-built Scanning Near-field Optical Microscope (SNOM) stage to
manipulate the emitter in the vicinity of the microsphere surface.
The recipe for the production of such probes is discussed by
K\"{u}hn \emph{et al}.~\cite{KuehnJM} while the alignment and
characterization of the microspheres and their WGMs are described
in Refs~\cite{GoetzingerAPB,Mazzei}. During all that follows, the
quality factor Q of the sphere could be measured by direct
spectroscopy using a narrow-band diode laser at $\lambda=670~nm$.
The fluorescent nanoparticle at the tip was excited through the
fiber, and a prism was used to extract part of the particle
emission that coupled to the WGMs~\cite{JOPTB2004}. Alternatively,
a microscope objective (NA=0.75) allowed us to collect both the
free space components of the nanoparticle fluorescence and the
scattering from the sphere~\cite{JOPTB2004}.

Figure~\ref{fig1}b shows part of the fluorescence spectrum
recorded via the prism when a bead (Red Fluorescent, Molecular
Probes, Inc.) of $500~nm$ in diameter was placed within a few
nanometers of the microsphere. The spectrum shows a FSR of
$0.85\,nm$ expected for a sphere used in this measurement of
$2R\approx 96~\mu m$ . By using polarized detection, we identified
the two dominant peaks in each FSR as TM and TE modes (see
Fig.~\ref{fig1}b). Since the small numerical aperture of our
detection path via the prism was optimized for coupling to low $n$
modes~\cite{GorodetskyOC}, we attribute these resonances to $n=1$
and the weaker ones to higher $n$ modes.

\begin{figure}[b!] \centering
\includegraphics[width=10 cm]{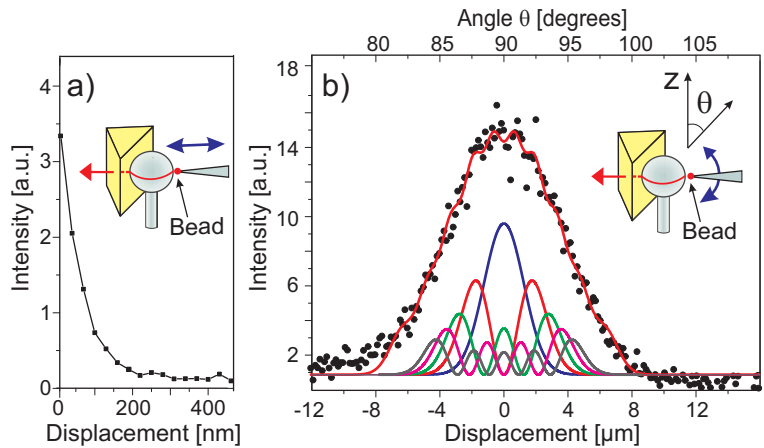}
\caption{ Total fluorescence intensity detected through the prism
coupler as a function of the sphere-bead distance (a) and the
particle's lateral position (b). The thick red line in b) is a fit
using the first ten WGMs. The weighted intensity distribution of
only the first five WGMs is plotted for clarity.} \label{fig2}
\end{figure}

Now, we examine the position dependence of the bead's coupling to
the WGMs. To do this, we tuned the emitter's angular coordinate
$\theta$ to the sphere equator~\cite{Mazzei} and then varied its
radial separation to the sphere surface using the SNOM distance
stabilization and scanning machinery (see insets in
Figs.~\ref{fig2}a and b). The fluorescence emitted into the WGMs
was collected through the prism coupler and detected by a
photomultiplier tube. Figure~\ref{fig2}a shows a characteristic
decrease expected for the evanescent coupling between the bead and
the WGMs of the microsphere. Next, we fixed the particle-sphere
separation to $5-10~nm$ and scanned the bead about the equator
along $\theta$. The symbols in Fig.~\ref{fig2}b show that the
fluorescence signal detected through the prism drops quickly
within $5 ^\circ$ or equivalently $4~\mu m$ about the equator. We
note that at room temperature the broad spectrum of a molecule
couples to many modes of different $m$. However, because WGMs with
higher $l-\left\vert m \right\vert$ values have larger mode
volumes and therefore lower electric fields at the equator, their
coupling efficiencies to both the bead fluorescence and to the
prism is reduced. The red curve in Fig.~\ref{fig2}b displays a fit
to the data, accounting for the contributions of the first ten
WGMs of different $m$. The profiles of the first five are plotted
under the experimental data where the blue curve represents the
fundamental mode, and the profile heights reflect the respective
weighting factors in the fit procedure. The data in
Fig.~\ref{fig2} demonstrate the local and controlled coupling of a
single nanoemitter to the WGMs of a microsphere.

\begin{figure}[b!] \centering
\includegraphics[width=10 cm]{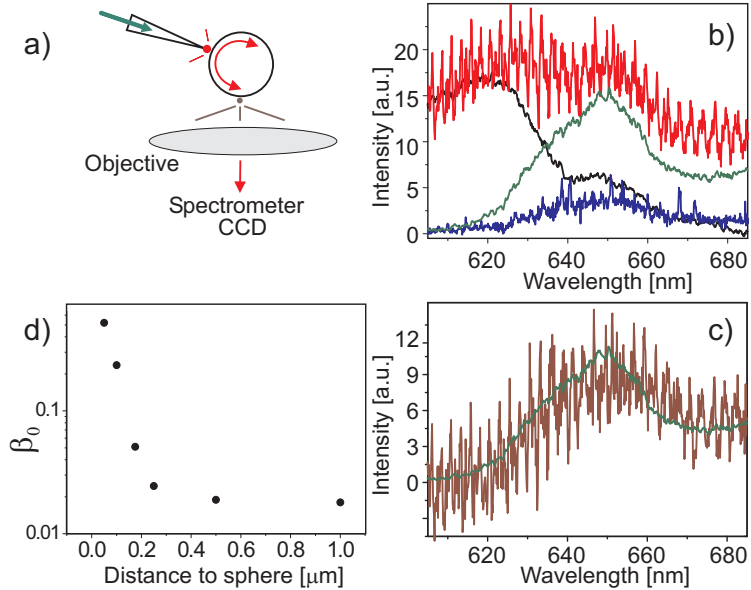}
\caption{a) Scheme of the cavity-mediated photon transfer
measurement. b) The black and the green curves display the
reference fluorescence spectra of the naked donor and the acceptor
beads, respectively. The red curve shows the recorded spectrum
when a donor is brought close to the sphere's surface. The blue
curve shows the same measurement after bleaching of the donor. c)
The brown curve plots the net emission spectrum of the acceptor as
a result of photon transfer. The green curve shows the normalized
spectrum of a free acceptor from (b) for comparison. d) Calculated
dependence of $\beta_0$ on emitter's separation from the
microsphere.} \label{fig3}
\end{figure}

Next, we discuss photon transfer between a donor and an acceptor
nanoparticle via the WGMs. In this experiment a sphere of
$2R=35\,\mu m$ was coated with a solution of acceptor beads
(Crimson, Molecular Probes, Inc.), $200\, nm$ in diameter. After
coating there were a total of less than 10 particles on the
surface of the microsphere and the under-coupled Q of the
fundamental mode was measured to be $3 \times 10^7$. Then we
retracted the prism to avoid losses due to output coupling and
imaged the location of the acceptor beads on the microsphere using
a CCD camera~\cite{JOPTB2004}. We located a single nanoparticle
close to the sphere equator and centered it in the confocal
detection path of the spectrometer (see Fig.~\ref{fig3}a). A
single donor bead (Red Fluorescent, Molecular Probes, Inc.) of
$500\, nm$ in diameter was attached to a tip and was excited
through the fiber with a laser power of $\approx 20~\mu W$. The
black and green curves in Fig.~\ref{fig3}b show reference
fluorescence spectra of the donor and acceptor beads,
respectively, recorded on a cover glass. Finally, we approached
the donor to the sphere and recorded the spectrum of the single
acceptor bead through the microscope objective.

The red curve in Fig.~\ref{fig3}b plots the spectrum obtained from
the location of the acceptor. The fast spectral modulations
provide a direct evidence of coupling to high-Q
WGMs~\cite{Chang_Campillo,Eversole,Yukawa_PRA_99,FanOPL2000}.
Comparison of this spectrum with the black and green spectra
reveals the coexistence of contributions from the donor and the
acceptor fluorescence. We remark that although our confocal
detection efficiently discriminates against light emitted at the
donor location, it is possible for this emission to couple to the
WGMs and get scattered into our collection path by the acceptor
bead. To take this into account, we subtracted the donor
fluorescence spectrum from the recorded (red) spectrum after
normalizing their short wavelength parts. Furthermore, to rule out
the possibility of direct excitation of the acceptor by the laser
light, we retracted the donor from the sphere, photobleached it
with an intense illumination of the excitation light and
approached it again to the sphere. The signal at the acceptor
position was then collected under exactly the same conditions and
is shown in Fig.~\ref{fig3}b. This contribution is clearly
negligible compared to the total emission (red curve), verifying
that the acceptor fluorescence has been almost entirely pumped by
the donor emission. After subtracting this small contribution, we
arrive at the brown curve in Fig.~\ref{fig3}c which coincides very
well with the fluorescence spectrum of the acceptor (also shown in
\ref{fig3}c for convenience). We note in passing that we have also
checked that bleaching the acceptor bead would result in the
disappearance of the longer wavelength part of the red spectrum in
Fig.~\ref{fig3}b. These measurements show, to our knowledge, the
first experimental realization of photon exchange between two
well-defined nanoemitters via shared high-Q modes of a
microresonator. Below, we discuss the underlying physical
phenomena.

Let us define the transfer efficiency $\eta_i$ as the probability
$\beta_i$ of the donor emitting a photon into the $i^{th}$ WGM and
subsequently for this photon to get absorbed by the acceptor. Then
the efficiency $\eta_i$ for a photon that is emitted by the donor
to be absorbed by the acceptor can be written as
\begin{equation}
\eta_i=\beta_{i}
\frac{\sigma_{A,abs}}{\sigma_{A,abs}+\sigma_{D,sca}+\sigma_{D,abs}+\sigma_{i,Q}},
\end{equation}
where the quotient stands for the probability of a cavity photon
being absorbed by the acceptor before getting lost in other
channels. Note that because the emission and absorption processes
are independent here (in contrast to ordinary FRET), a simple
multiplication of probabilities is appropriate~\cite{LeungJCP88}.
The parameter $\sigma_{A,abs}$ denotes the absorption cross
section of the acceptor whereas cross sections $\sigma_{D,sca}$
and $\sigma_{D,abs}$ quantify losses out of the mode due to
scattering and absorption of a photon by the donor. Finally,
$\sigma_{i,Q}$ is a cross section signifying all losses associated
with the measured Q of the microsphere, including those caused by
scattering from the acceptor. A small mode volume is therefore,
important for the enhanced emission rate of photons from the donor
into the sphere and enters the quantity $\beta_i$. Furthermore,
the high Q and hence small $\sigma_{i,Q}$ result in the enhanced
absorption probability of photons by the acceptor. Both effects
are absent if light is transferred merely by a waveguide or
optical fiber.

The cross sections used in Eq.~(1) are, strictly speaking, defined
for evanescent illumination and different from those commonly
quoted for plane wave excitation. The deviation between the two
quantities can be notable, but it has been shown that it remains
well within a factor of 3 even for strongly scattering silver
particles of diameter 200~nm at plasmon
resonance~\cite{Quinten:99}. Thus, in our case it is appropriate
to use the conventional values of the cross sections for obtaining
an order of magnitude estimate. The absorption cross section of
the acceptor particle can be taken as $\sigma_{A,abs} \approx
10^{-11}~cm^2$, assuming $\sigma_{abs} \approx 10^{-16}~cm^2$ per
molecule~\cite{Schaefer} and $10^5$ molecules per particle. Since
due to the Stokes shift of molecular fluorescence the donor does
not absorb its own emission very efficiently, we can neglect
$\sigma_{D,abs}$. In addition, we obtain $\sigma_Q = 2\times
10^{-12}~cm^2$ for a fundamental mode based on $Q=3 \times 10^7$.
Since in our experiment the measured $Q$ remained unchanged as the
tip approached the microsphere, we conclude that $\sigma_{D,sca}$
was negligible compared to $\sigma_Q$~\cite{GoetzingerOL}. We
find, therefore, that for a fundamental mode, the quotient in Eq.
(1) is about $10^{-4}$ considering a single molecule acceptor.

In the weak coupling regime, the spontaneous emission
rate $\Gamma$ can be written as $\Gamma =\frac{2\pi }{\hbar ^{2}}\left\vert \left\langle e\left\vert \mathbf{%
E}.\mathbf{D}\right\vert g\right\rangle \right\vert ^{2}\rho
(\omega )$ where \textbf{E} is the fluctuating vacuum field at the
location of the emitter, \textbf{D} is the dipole operator
associated with the optical transition at hand, and $\rho$ is the
density of photon states. Hence, the strength of emission into
different WGMs and consequently $\beta_i$, are proportional to the
projection of each mode's field intensity $\left\vert E_i
\right\vert^2$ at the sphere surface onto the dipolar axis. In
what follows we calculate $\beta_i$ for the fundamental mode and
use its scaling with $\left\vert E_i \right\vert^2$ to evaluate
the contribution of other WGMs to the energy transfer process.

Since the seminal work of Purcell, it is known that spontaneous
emission of a narrow-band dipole is enhanced when it is coupled to
a resonator mode~\cite{Berman_book_1996}. This enhancement is
reduced if the linewidth of the dipole is broadened to $\Gamma_b$,
greater than the cavity linewidth $\Gamma_{cav}$, as is the case
for $\Gamma_{cav} = 6 \times 10^{-5} \, nm$ and $\Gamma_b \approx
20 \, nm$ in our experiment. The ratio $\beta$, of the emission
into the cavity mode to the total emitted power is thus, given by
$\beta\approx \beta_0 (\Gamma_{cav}/\Gamma_b)$~\cite{Ujihara}
whereby $\beta_0$ represents the fraction for a narrow-band
emitter. Note that since $\beta_0 \propto Q$, in this case $\beta$
becomes independent of $Q$. To calculate $\beta_0$ for a
fundamental mode, we calculated the power radiated into this
mode~\cite{Klimov:96} by approximating the emission of a randomly
oriented dipole by a spherical wave and calculating its overlap
with the mode~\cite{Barton-88}. The finite Q of the sphere was
accounted for by using a complex $N$~\cite{Dung-01}. Figure 3d
shows the result as a function of the distance between the emitter
and the sphere's surface. We find that $\beta_0= 0.5$ at a
distance of 50~nm from the sphere surface, leading to $\beta_i=
1.5 \times 10^{-6}$.

For other WGMs, higher $n$ and $l-\left\vert m \right\vert$ result
in an increase of the mode volume and lower $\left\vert E_i
\right\vert^2$ values. By computing the mode functions of the
various WGMs, we have determined $\left\vert E_i \right\vert^2$ on
the sphere surface normalized to its value for the fundamental
mode. Furthermore, by calculating the diffraction limited Q for
various $n$~\cite{Datsyuk:92}, we have determined the dependence
of the quotient in Eq. (1) on this parameter. Combining these
results we find that for each $l$ the contribution to
$\eta=\sum{\eta_i}$ of modes with $n> 10$ drops by an order of
magnitude. We also find that the first 40 modes with different
$l-\left\vert m \right\vert$ values account for half the
contribution of all modes. Considering that the fluorescence
spectrum of a bead spans about 20 FSRs ($FSR= 2.3~nm$ for
$2R=35~\mu m$) and taking into account both TE and TM modes, we
conclude that $\eta$ is about $10 \times 40 \times 20 \times 2
\approx 2 \times 10^4$ times larger than that of a single
fundamental mode. Putting all the above-mentioned information
together, we find that $\eta \approx (1.5 \times 10^{-6}) \times
10^{-4} \times 2 \times 10^4 \approx 10^{-6}$ for a single
molecule acceptor which is more than six orders of magnitude
larger than the free-space rate for absorbing a photon emitted at
a distance of about 35 micrometers.

We have demonstrated that the application of scanning probe
techniques allows one to achieve an on-command and efficient
evanescent coupling between a nanoscopic emitter and a WGM
resonator. In the current system we have coupled the broad-band
emission of dye molecules to a large number of resonator modes in
a dissipative energy transfer process. Under these conditions, the
role of the high cavity Q in the enhancement of spontaneous
emission has been negligible. Instead, the importance of the high
Q has been to circulate each photon a large number of times,
increasing its chance of interaction with the acceptor. However,
at low temperatures emitter transitions as narrow as a few tens of
MHz can be achieved, enhancing the photon transfer efficiency
between single emitters by an additional five orders of magnitude.
Furthermore, one could get around the inhomogeneous distribution
of resonance frequencies by local application of electric
fields~\cite{Hettich:02} and use our experimental arrangement to
achieve controlled coherent coupling of individual quantum
emitters mediated by a high-Q microcavity~\cite{Imamoglu}. Indeed,
cryogenic efforts have already demonstrated the coupling of single
molecules to WGMs~\cite{Norris-97} and their manipulation at the
end of a tip~\cite{michaelNat}.

We acknowledge funding from the DFG (SP113) and the Swiss National
Foundation. L. de S. Menezes acknowledges the fellowship from the
Alexander von Humboldt Stiftung. A. Mazzei acknowledges financial
support of NaF\"{o}G, Berlin.

\end{document}